\documentstyle[prb,aps,epsf]{revtex}
\begin{document}

\title{Influence of exciton-phonon interaction on long energy transport in J~-~aggregates.}
\author{E.A. Bartnik\thanks{bartnik@fuw.edu.pl}, M. Bednarz\thanks{bednarz@fuw.edu.pl}}
\address{Institute of Theoretical Physics, Warsaw University, Ho\.{z}a Street 69, 00-681 Warsaw, Poland}
\maketitle

\begin{abstract}
This paper presents a theoretical model intended to address the question of energy transfer in two-dimensional molecular assemblies such as Scheibe aggregates. A new phonon-exciton interaction is introduced to explain the exciton width in J aggregates. It is shown that the long range energy transport can occur for weakly interacting acceptors.
\end{abstract}

\pacs{}

\section{Introduction}
Compact and regular two dimensional arrangements of oxycyanine dye bound to hydrophobic carbon tails show remarkably narrow absorption and fluorescence peaks [1-5]. In these so called J-aggregates it is believed that the excitation of a dye molecule becomes delocalized (exciton) due to the strong dipole-dipole interactions between delocalized electrons in adjacent dye molecules. It was also shown [3-5] that these structures can support practically loss-free energy transport over large distances. Molecules of a suitable acceptor (a molecule similar to that used in J-aggregate, but with thiocyanine instead of oxycyanine) were able to absorb almost all the energy, even at concentrations as low as 1:10 000. It is, however, astonishing,  that energy transport is more efficient when the coupling to the acceptor is weak [5]. Quantum mechanical mechanism of this effect was developed in [7,8] in terms of exciton trapping by shallow bound state of large spatial extend. However, the theory assumed 'ideal' excitons, which would imply extremely small absorption width in J-aggregate. Clearly a dissipation mechanism is needed to explain the observed experimentally absorption width, which, moreover, should not destroy the delicate exciton trapping needed to explain long range energy transport. Investigation of possible dissipation mechanisms is the aim of this paper. It will be shown that the usually assumed [6] exciton-phonon coupling is not up to the task and that only a mechanism destroying excitons will do. We propose, therefore, a new phonon-exciton interaction, which has the required properties and, moreover, leads to a solvable model. Then we demonstrate that our interaction permits a long range energy transfer in two dimensional structures.

\section{Ideal excitons coupled to photons}
We start with the one dimensional Hamiltonian for the excitons [6,7]. We have

	\begin{equation}
\hat{H}=\sum_{n=-\infty}^{\infty}[\Omega A_{n}^{+}A_{n}-JA_{n}^{+}(A_{n-1}+A_{n+1})],
	\end{equation}
where $A^{+}_{n}$, $A_{n}$ are the creation and annihilation operators, respectively, of an excitation at site n, $\Omega$ is the excitation energy of a single dye molecule and J is the resonant interaction energy between adjacent sites. We use one dimensional formalism for simplicity's sake, two dimensional generalization is immediate. Hamiltonian (1) is diagonalized in momentum space. Introducing Bloch waves with quasi momentum k we have

	\begin{eqnarray}
A_{n}&=&\frac{1}{\sqrt{2\pi}}\int_{-\pi}^{\pi}dke^{ikn}A_{k},\\
A_{k}&=&\sum_{n=-\infty}^{+\infty} e^{-ikn}A_{n}.
	\end{eqnarray}

Our Hamiltonian  (1) is now

	\begin{equation}
\hat{H}=\int_{-\pi}^{\pi}\Omega(k)A_{k}^{+}A_{k}dk
	\end{equation}
with the dispersion relation

	\begin{equation}
\Omega(k)=\Omega-2J\cos(k).
	\end{equation}

Again for simplicity we use scalar photons with the coupling to the dye within rotating wave approximation [8], so the Hamiltonian for the exciton-photon system is

\begin{eqnarray}
\hat{H}&=&\int dk\Omega(k)A^{+}_{k}A_{k} + \int d^{3}p\varepsilon(p)c_{p}^{+}c_{p} + \\
& &\int d^{3}pdkg(p)\delta(p_{||}-k)(A_{k}c^{+}_{p} + c_{p}A^{+}_{k}),\nonumber
\end{eqnarray}
where the second integral is the energy of the photons with the usual dispersion relation

\begin{equation}
\varepsilon(p)=c|p|
\end{equation}
and the last term is the photon-exciton interaction. Note that the sum of exciton and phonon numbers is conserved (by the virtue of rotating wave approximation) and that we have delta of momentum conservation along the chain. At the momenta of interest the molecular formfactor $g(p)$ is practically a constant, which we denote $g_{0}$, and its momentum dependence is of importance in loop integrals only: it gives the natural line width of an atomic state, much smaller than the experimentally observed width [8]. It is important to realize that since the wavelength of visible light is of the order 1000~\AA, this corresponds to momenta of the order $10^{-3}$ $\AA^{-1}$, so to a good approximation light produces excitons with negligible momentum, i.e. from the bottom of the band. In this form the model is analytically solvable. Working within one photon-one exciton sector we use for the wave function the ansatz

	\begin{equation}
|\psi>=\int dk\varphi(k)A_{k}^{+}|0> + \int d^{3}{p}\psi_{p_{0}}(p)
c_{p}^{+}|0>,
	\end{equation}
where $|0>$ is the Fock vacuum (which is an exact eigenstate of the theory). This leads to a system of equations for the unknown functions $\varphi(k)$ and $\psi_{p_{0}}(p)$

\begin{equation}
\left\{
\begin{array}{lll}
z\varphi(k)&=&\Omega(k)\varphi(k)+\int d^{3}pg(p)\delta(p_{||}-k)\psi_{p_{0}}(p)\\
z\psi_{p_{0}}(p)&=&\varepsilon(p)\psi_{p_{0}}(p) + g(p)\varphi(p_{||}).
\end{array}\right.
\end{equation}
where the photon wavefunction $\psi_{p_{0}}(p)$ is a sum of plane and scattered wave

\begin{equation}
\psi_{p_{0}}(p)=\delta_{3}(p-p_{0})+\Psi_{p_{0}}^{+}(p),
\end{equation}
where $p_{0}$ is the momentum of the incoming light. System (9) is easily solved and we get for the experimentally observed photon wavefunction

\begin{equation}
\Psi^{+}_{p_{0}}(p)=\frac{\delta(p_{0||}-p_{||})}{(\varepsilon(p_{0})-\varepsilon(p)}
\frac{g(p)g(p_{0})}{(z-\Omega(p_{||})-G(p_{||}))}.
\end{equation}

As we see, the scattered wave has a cylindrical outgoing wave (note the one dimensional delta function) multiplied by scattering amplitude. The absorption intensity is proportional therefore to

\begin{equation}
\left|\frac{g(p)g(p_{0})}{z-\Omega(p_{||})-G(p_{||})}\right|^{2}
\end{equation}
where we have introduced the notation

\begin{equation}
G(k)=\int d^{3}p\frac{g^{2}(p)\delta(p_{||}-k)}
{z-\varepsilon(p)-i\varepsilon},
\end{equation}
for small $k$ (our case) the real part of $G(k)$ is negligible, while the imaginary part is

\begin{equation}
G(0)=i\pi g_{0}^{2}.
\end{equation}

The absorption probability has therefore the familiar lorentzian shape

\begin{equation}
\frac{1}{(z-\Omega(0))^{2}+\pi^{2}g_{0}^{4}}
.\end{equation}

Due to the smallness of $g_{0}$ [8], the absorption peak has a finite, albeit very small with. In order to get a more realistic value, we need a dissipation mechanism, i.e. interaction with phonons.

\section{Phonon-exciton interaction}
Like everybody else we assume that excitons interact with phonons, since the J-aggregate structure is not very stiff. We take the phonon Hamiltonian in the familiar form

\begin{equation}
\hat{H}=\int_{-\pi}^{\pi}\omega(q)a^{+}_{q}a_{q}dq.
\end{equation}

where $a^{+}_{q}$, $a_{q}$ are the (bosonic) creation and annihilation operators for phonons with the dispersion relation

	\begin{equation}
		\begin{array}{rclcl}
\omega(q)&=&\sqrt{2\omega_{0}(1-\cos(q))}&\cong&\sqrt{\omega_{0}}|q|,
		\end{array}
	\end{equation}

One usually assumes a trilinear coupling exciton-phonon [6]

		\begin{equation}
\hat{H_{I}}=\int\chi(k,q)A_{k+q}^{+}A_{k}(a_{q} + a^{+}_{-q})dkdq
		.\end{equation}

However, this coupling does not change the exciton number: This interaction can only push excitons to the bottom of the band, but never destroy them! Since excitons are created with negligible momentum anyway, this mechanism, although certainly present, cannot contribute significantly to the observed light absorption width. We need therefore an interaction which will explicitly destroy excitons, regardless their energy. The simplest possibility is the annihilation of an exciton into  2 phonons. After all, the excited dye molecule contains a delocalized electron, which can not only reradiate light, but also give up its energy to the vibrations of the whole structure. We of course realize that it is difficult for two phonons to carry away 2-3~eV of energy, so we call them 'effective'. We therefore propose the interaction in the form

	\begin{equation}
\hat{H_{I}}=\int\chi(k,q)(A_{k}a_{k-q}^{+}a^{+}_{q}+A_{k}^{+}a_{k-q}a_{q})dkdq.
	\end{equation}
where $\chi(k,q)$ is the formfactor, which, for the sake of simplicity, we assume is constant. Our interaction has the great advantage that it leads to a solvable model. Our excitons coupled to photons and phonons are described by a free part

	\begin{equation}
\hat{H_{0}}=\int\Omega(k)A_{k}^{+}A_{k}dk + \int\omega(q)a_{q}^{+}a_{q}dq + \int\epsilon(p)c_{p}^{+}c_{p}d^{3}p
	\end{equation}
and the interactions

	\begin{eqnarray}
\hat{H_{I}}&=&\int\chi(A_{k}a_{k-q}^{+}a^{+}_{q}+A_{k}^{+}a_{k-q}a_{q})dkdq +\nonumber\\
 & &\int g_{p}(A_{k}c^{+}_{p}+c_{p}A_{k}^{+})\delta(k-p_{||})dkd^{3}p.
	\end{eqnarray}
One exciton sector is solved by the ansatz

	\begin{eqnarray}
|\Psi>&=&\int dk\varphi(k)A_{k}^{+}|0>+\int dq_{1}dq_{2}\Psi_{ph}(q_{1},q_{2})a_{q_{1}}^{+}a^{+}_{q_{2}}|0>+\nonumber\\
& &\int d^{3}p\Psi_{p_{0}}(p)c^{+}_{p}|0>
	\end{eqnarray}
which leads to a coupled set of equations

\begin{equation}
\left\{
\begin{array}{lll}
z\varphi(k)&=&\Omega(k)\varphi(k)+2\int dq\chi(q)\psi_{ph}(q,k-q)+\\
 & &\int d^{3}pg_{p}\psi_{p_{0}}(p)\delta(p_{||}-k)\\
z\psi_{p_{0}}(p)&=&\varepsilon(p)\psi_{p_{0}}(p)+g_{p}\varphi(p_{||})\\
z\psi_{ph}(q_{1},q_{2})&=&\chi(q_{2})\varphi(q_{1}+q_{2})+(\omega(q_{1})
+\omega(q_{2}))\psi_{ph}(q_{1},q_{2}).
\end{array}\right.
\end{equation}
where the energy is

\begin{equation}
z=\varepsilon(p_{0})
\end{equation}

The scattered photon wave function can be written in the form

	\begin{equation}
\Psi_{p_{0}}^{+}(p)=\frac{g_{p_{0}}g_{p}\delta(p_{0||}-p_{||})}{\{z-\Omega(p_{||})-\Gamma_{z}(p_{||})-G(p_{||})\}\{\varepsilon(p_{0})-\varepsilon(p)\}}
	,\end{equation}
where we have introduced the function

	\begin{equation}
\Gamma_{z}(k)=2\int_{-\pi}^{\pi}\frac{\chi^{2}(q)}{z-\omega(q)-\omega(k-q)-i\varepsilon}dq
	.\end{equation}

Absorption cross section is therefore proportional to

	\begin{equation}
\left|\frac{1}{z-\Omega(k)-\Gamma_{z}(k)}\right|^{2}
	.\end{equation}
Remembering that excitons are created with zero momentum we need the function (26) at k=0, we have

\begin{equation}
\Gamma_{z}(0)=2\chi^{2}\int_{-\pi}^{\pi}\frac{dq}{z-2\omega(q)-i\varepsilon}
.\end{equation}
Assuming linear dispersion for the phonons (cf. eq. 16) we can perform the integration explicitly

	\begin{equation}
\frac{1}{(z-\Omega(0)+\frac{2\chi^{2}}{\sqrt{\omega_{0}}}\ln|1-\frac{2\pi\sqrt{\omega_{0}}}{z}|)^{2}+
\frac{4\pi^{2}\chi^{4}}{\omega_{0}}}
	.\end{equation}

The logarithm does not play, numerically speaking, any significant role, therefore we have obtained a familiar lorentzian shape. Its width is given by

\begin{equation}
Im^{2}\Gamma_{z}(0)=\frac{4\pi^{2}\chi^{4}}{\omega_{0}}
\end{equation}
and depends on $\chi$ and $\omega_{0}$. Since experimentally it is of the order of 0.026~eV, we can estimate $\chi$ if we assume $\omega_{0}$, which we arbitrally have taken to be 4.0~eV. The $\Omega(0)$ parameter gives the position of the absorption peak, taking it to be 3.104~eV we get an excellent agreement with the experiment.

\section{Acceptor in the presence of phonons}

    Presence of an acceptor molecule, which we place at site $n=0$, introduces an additional term in the exciton Hamiltonian, which now reads

\begin{equation}
\hat{H}=\sum_{n=-\infty}^{\infty}[\Omega A_{n}^{+}A_{n}-JA_{n}^{+}(A_{n-1}+A_{n+1})]-(\Omega-\Omega_{0})A_{0}^{+}A_{0}
	,\end{equation}
where $\Omega_{0}$ is the acceptor excitation energy [7]. In momentum space this is equivalent to the introduction of a separable potential with constant formfactor $f(k)$

\begin{equation}
\hat{H}=\int\Omega(k)A_{k}^{+}A_{k}dk + \lambda \int dkf(k)A^{+}_{k}\int d\tilde{k}f(\tilde{k})A_{\tilde{k}}
\end{equation}
and

\begin{equation}
\lambda=\frac{\Omega_{0}-\Omega}{2\pi}
.\end{equation}

As shown in [7,8] for small values of lambda, leading to a shallow bound state, we get large absorption, while paradoxically large lambda does induce strong absorption. The question now is: will the interaction with phonons destroy this delicate mechanism? A satisfying feature of our interaction is that the model is still analytically solvable. The full interaction is now

	\begin{eqnarray}
\hat{H_{I}}&=&\int\chi(q)(A_{k}a_{k-q}^{+}a^{+}_{q}+A_{k}^{+}a_{k-q}a_{q})dkdq + \nonumber \\
 & &\int g_{p}(A_{k}c^{+}_{p}+c_{p}A_{k}^{+})\delta(k-p_{||})dkd^{3}p + \nonumber \\
 & &\lambda \int dkf(k)A^{+}_{k}\int d\tilde{k}f(\tilde{k})A_{\tilde{k}}
	,\end{eqnarray}
which gives now a set of equations in one exciton - one photon - two phonons sectors (which are coupled)

\begin{equation}
\left\{
\begin{array}{lll}
z\varphi(k)&=&\Omega(k)\varphi(k)+2\int dq\chi(q)\psi_{ph}(q,k-q)+\\
 & &\lambda f(k)\int d\tilde{k}f(\tilde{k})\varphi(\tilde{k})
+\int d^{3}pg_{p}\psi_{p_{0}}(p)\delta(p_{||}-k)\\
z\psi_{p_{0}}(p)&=&\varepsilon(p)\psi_{p_{0}}(p)+g_{p}\varphi(p_{||})\\
z\psi_{ph}(q_{1},q_{2})&=&\chi(q_{2})\varphi(q_{1}+q_{2})+(\omega(q_{1})
+\omega(q_{2}))\psi_{ph}(q_{1},q_{2}).
\end{array}\right.
\end{equation}

Again the system can be solved and we get the scattering part of the photon wave function

\begin{equation}
\Psi_{p_{0}}^{+}(p)=\frac{g_{p_{0}}g_{p}\delta(p_{0||}-p_{||})}
{\{z-\Omega(p_{||})-\Gamma_{z}(p_{||})-G(p_{||})\}\{\varepsilon(p_{0})-\varepsilon(p)\}}+
\end{equation}
\begin{displaymath}
\frac{g_{p}g_{p_{0}}f^{2}(p_{||})}
{
\{\varepsilon(p_{0})\!-\!\varepsilon(p)\}\{z\!-\!\Omega(p_{||})\!-\!\Gamma_{z}(p_{||})\!-\!G(p_{||})\}
\{z\!-\!\Omega(p_{0_{||}})\!-\!\Gamma_{z}(p_{0_{||}})\!-\!G(p_{0_{||}})\}
}/
\end{displaymath}
\begin{equation}
\{\frac{1}{\lambda}-\int\frac{d\tilde{k}f^{2}(\tilde{k})}
{z-\Omega(\tilde{k})-\Gamma_{z}(\tilde{k})-G(\tilde{k})}\}
\end{equation}
showing the effects of acceptor and phonon interactions in all its gory details. The last line is the effect of the acceptor. We will obtain a significant enhancement if  this term is small. As assumed above we can take

	\begin{equation}
\Gamma_{z}(k)\simeq\Gamma_{z}(0)= i \delta_{exp}
	\end{equation}
where $\delta_{exp}$ is the experimental width. The whole behavior is therefore controlled by the integral

\begin{equation}
\Sigma(z)=\int_{-\pi}^{\pi}\frac{dk}{z-\Omega(k)-i\delta_{exp}}
.\end{equation}

For suitable values of coupling lambda real part of $\Sigma$ can be canceled, leading to a large enhancement, provided that the imaginary part is reasonably small. Inspection of FIG.~1

\begin{figure}[htb]
\centerline{
\epsfysize=180pt
\epsffile{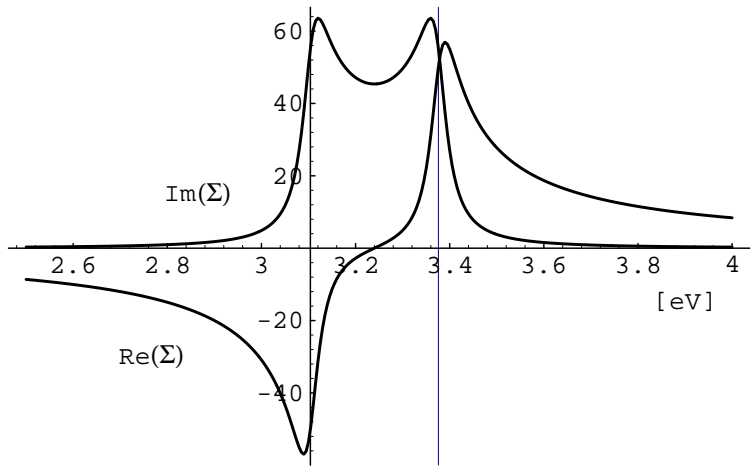}
}
\caption{Function $\Sigma(z)$, 1 dim case.}
\end{figure}

, where the y axis is at the threshold for exciton production, shows indeed that large values of $\frac{1}{\lambda}$ (i.e. lambda small!) are favored. Absorption spectra with $\frac{1}{\lambda}$ small (FIG.~2) and $\frac{1}{\lambda}$ (FIG.~3) show an enhancement of the order of $50 \%$.

\begin{figure}[htb]
\centerline{
\epsfysize=180pt
\epsffile{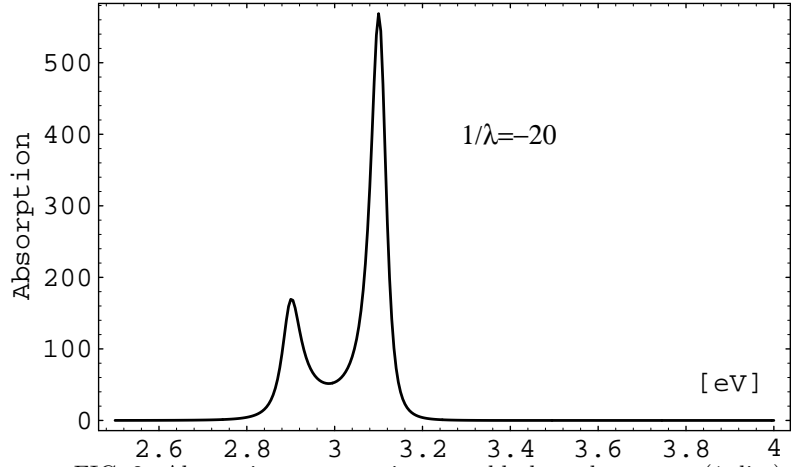}
}
\caption{Absorption cross section - weakly bound acceptor (1 dim)}
\end{figure}

\begin{figure}[htb]
\centerline{
\epsfysize=180pt
\epsffile{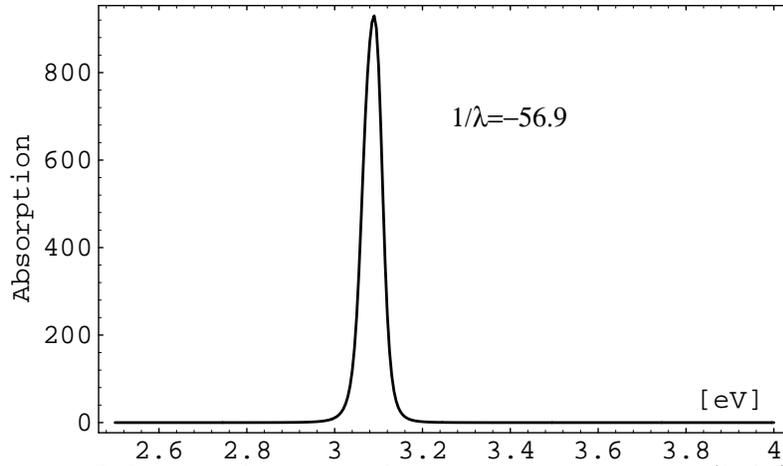}
}
\caption{Absorption cross section - strongly bound acceptor (1 dim)}
\end{figure}

The effect is there, but disappointingly small. The reason lies in the large value of the imaginary part of sigma (cf. FIG.~1).
We must notice however that this is a one dimensional model. In two dimensions we would get formally the same expressions, but all integrations are 2 dimensional. In particular the function sigma, after angular integration would be

	\begin{equation}
\Sigma(z)=\int_{0}^{\pi}\frac{kdk}{z-\Omega(k)-i\delta_{dos}}
	\end{equation}
and has integration measure $kdk$, which suppresses small $k$ - just what is needed. The imaginary part at threshold is now strongly reduced (FIG.~4).

\begin{figure}[htb]
\centerline{
\epsfysize=180pt
\epsffile{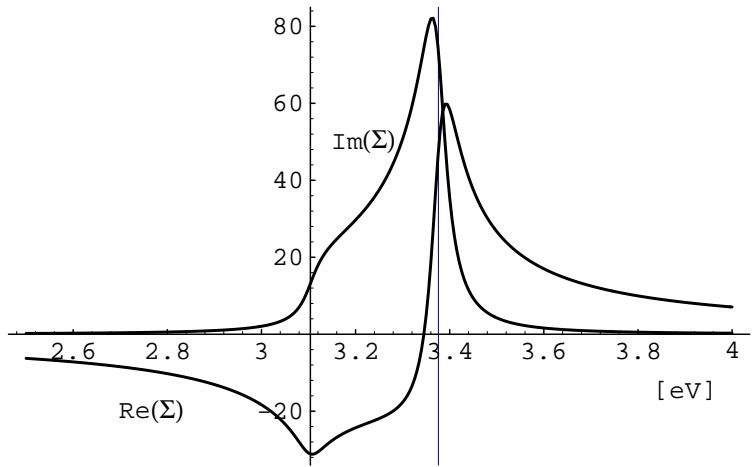}
}
\caption{Function $\Sigma(z)$, 2 dim case.}
\end{figure}

The absorption enhancement at small lambda now is of the order of $700 \%$ (FIG.~5,6).
\begin{figure}[htb]
\centerline{
\epsfysize=180pt
\epsffile{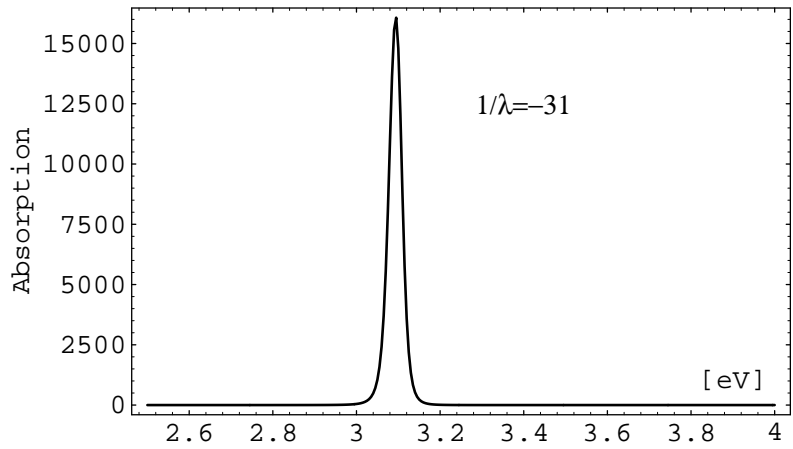}
}
\caption{Absorption cross section - weakly bound acceptor (2 dim)}
\end{figure}

\begin{figure}[htb]
\centerline{
\epsfysize=180pt
\epsffile{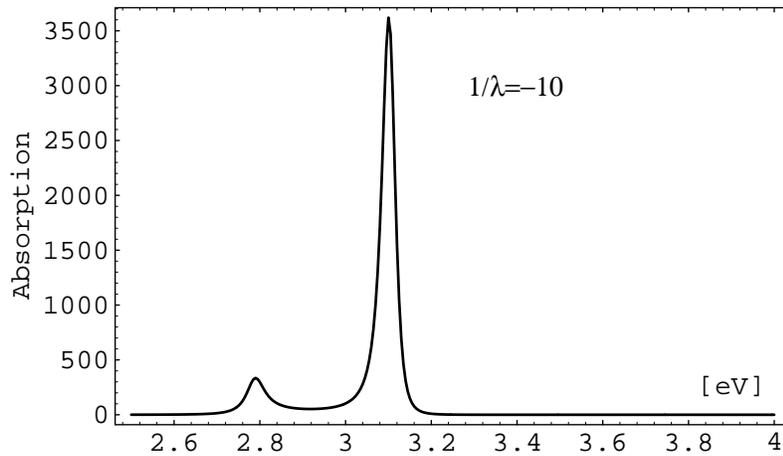}
}
\caption{Absorption cross section - strongly bound acceptor (2 dim)}
\end{figure}

It is seen that the fact that J aggregate is a two dimensional structure is of paramount importance.

\section{Conclusions and outlook}
    We have shown that our postulated mechanism: exciton annihilation into 2 phonons leads to a solvable model. It describes correctly the width of J-aggregate absorption spectra. In the case of introduced acceptors, the large absorption on shallow bound states is confirmed, but two dimensional geometry seems essential. Of course full two dimensional calculation is needed and will be presented shortly. In order to describe fluorescence spectra, we have to include the usual energy loss by phonon radiation mechanism [6]. In conclusion let us remark that absorption energy on shallow bound states, postulated already in [7] not only survives the introduction of realistic dissipation mechanisms, but acquires new features allowing for a better agreement with experiment.

\end{document}